\documentclass[structabstract]{aa}
\usepackage{txfonts}
\usepackage{graphicx,amssymb}
\usepackage{array}
\usepackage{natbib}
\usepackage{multirow}

\begin{document}

\title{Star-galaxy separation in the AKARI NEP Deep Field}

\author{A.~Solarz\inst{1,2}
  \and A.~Pollo\inst{2,3,4}
\and T.~T.~Takeuchi\inst{1,5}
\and A.~P\c epiak\inst{2}
\and H.~Matsuhara\inst{6}
\and T.~Wada\inst{6}
\and S.~Oyabu\inst{6}
\and T.~Takagi\inst{6}
\and T.~Goto\inst {7,8}
\and Y.~Ohyama\inst{9}
\and C.~P.~Pearson\inst{10,11,12}
\and H.~Hanami\inst{13}
\and T.~Ishigaki\inst{13}
}

\institute{Department of Particle and Astrophysical Science, Nagoya University, Furo-cho, Chikusa-ku, Nagoya 464-8602, JAPAN\\
         \email{solarz.aleksandra.alicja@c.mbox.nagoya-u.ac.jp}
         \and
         The Astronomical Observatory of the Jagiellonian University, ul.\ Orla 171, 30-244 Krak\'{o}w, POLAND
         \and
         Center for Theoretical Physics of the Polish Academy of Sciences, al.\ Lotnik\'{o}w, 32/46, 02-668, Warsaw, POLAND
         \and
         The Andrzej So{\l}tan Institute for Nuclear Studies, ul.\ Ho\.{z}a 69, 00-681 Warsaw, POLAND
         \and
         Institute for Advanced Research, Nagoya University, Furo-cho, Chikusa-ku, Nagoya 464-8601, JAPAN
         \and
         Institute of Space and Astronautical Science, Japan Aerospace Exploration Agency, Sagamihara, Kanagawa 252-5210, JAPAN
	\and
	Institute for Astronomy, University of Hawaii, 2680 Woodlawn Drive, Honolulu, HI, 96822, USA
	\and
	National Astronomical Observatory, 2-21-1 Osawa, Mitaka, Tokyo, 181-8588, Japan
	\and
	Academia Sinica, Institute of Astronomy and Astrophysics, Taiwan
	\and
	Space Science and Technology Department, CCLRC Rutherford Appleton Laboratory, Chilton, Didcot, Oxfordshire OX11 0QX, UK
	\and
	Department of Physics, University of Lethbridge, 4401 University Drive, Lethbridge, Alberta T1J 1B1, Canada
	\and
	Astrophysics Group, Department of Physics, The Open University, Milton Keynes, MK7 6AA, UK
	\and
	Physics Section, Faculty of Humanities and Social Sciences, Iwate University, Morioka 020-8550, Japan
}

\date{Received <date>/ Accepted <date>}

\abstract {It is crucial to 
develop a method for classifying objects detected in deep surveys at infrared wavelengths. We specifically need a method to separate 
galaxies from stars using only the infrared information to study the properties of 
galaxies, e.g., to estimate the angular correlation function, without introducing any additional bias.} 
{We aim to separate stars and galaxies in the data from the AKARI North Ecliptic Pole (NEP) Deep survey collected in nine AKARI / IRC bands from 2 to 24 $\mu$m that cover the near- and mid-infrared wavelengths (hereafter NIR and MIR). 
We plan to estimate the correlation function for NIR and MIR galaxies from a sample selected according to our criteria in future research.} 
{ We used support vector machines (SVM) to study the distribution of stars and galaxies in the AKARIs multicolor space. We defined the training samples of these objects by calculating their infrared stellarity parameter ($sgc$). We created the most efficient classifier and then tested it on the whole sample. 
We confirmed the developed separation 
with auxiliary optical data obtained by the Subaru telescope and by creating Euclidean normalized number count plots.}
{ We obtain a 90\% accuracy in pinpointing galaxies and 98\% accuracy for stars in infrared multicolor space with the infrared SVM classifier. The source counts and comparison with the optical data (with a consistency of 65\% for selecting stars and 96\% for galaxies) confirm that our star/galaxy separation methods are reliable.}
{ The infrared classifier derived with the SVM method based on infrared $sgc-$ selected training samples proves to be very efficient and accurate in selecting stars and galaxies in deep surveys at infrared wavelengths carried out without any previous target object selection.}

\maketitle 

\section{Introduction}

The first proof that various types of extragalactic sources evolved with cosmic epochs was delivered in 1950s by surveys of extragalactic radio sources and quasars, which revealed an excess of faint sources when compared with uniform distribution models (e.g. \citealt{ryle}; \citealt{radio}).
Interest in studying the deep Universe became much greater after the discovery of the excess of faint blue galaxies in optical passbands with photographic plates (e.g. \citealt{kron}; \citealt{hdf}; \citealt{ellis}).
This revelation was followed by discoveries of excess numbers of faint sources at early cosmic epochs in all wavelengths: in X-rays by the ROSAT X-ray Observatory (e.g. \citealt{rosat}), Chandra (e.g. \citealt{chan}), and XMM-Newton (e.g. \citealt{xmm}); in mid- and far- infrared by ISO (e.g. \citealt{elais}; \citealt{firback}; \citealt{lh}; \citealt{takeuchi}), IRAS (e.g. \citealt{iras}; \citealt{bertin}), and later by the Spitzer Space Telescope (e.g. \citealt{spitzer}; \citealt{dole}; \citealt{frayer}), and, when observational techniques became available, in the submillimeter range (e.g. \citealt{submm}; \citealt{hers}; \citealt{oliv}; \citealt{val}).
This was the first step toward modern studies of evolutionary processes.

The theoretical motivation for studying cosmic evolution arose from observations of the local Universe, which show a very diverse galaxy distribution, whereas the early Universe was almost uniform (e.g. \citealt{pee}). Those complex patterns are the result of tiny density fluctuations that interacted and increased gravitationally as the Universe expanded. Galaxy distribution can be studied in various statistical ways. 
The recent cosmological probes provide more and more proof that the large-scale structure of the Universe was created according to the hierarchical 
formation scenario.
 This describes the formation and evolution of galaxies inside halos of dark matter, which interacted gravitationally, resulting in their growth 
through mergers \citep{wr}.
Today the clustering process of dark matter halos 
is adequately understood (e.g. \citealt{mowhite}, \citealt{gao}), but this is not the case for the clustering of galaxies; 
establishing the link between dark matter halos and the baryonic component is one of 
the most challenging tasks modern cosmology has to deal with. 

The AKARI satellite (previously known as ASTRO-F or IRIS - InfraRed Imaging 
Surveyor) was designed to carry out infrared observations with a sensitivity and resolution higher than preceding missions.
It was launched by JAXA's MV8 vehicle on February 22, 2006, and, among many others, it performed a deep survey of the North Ecliptic Pole region (hereafter NEP), which we aim to use to explore the mid-infrared properties of galaxies, in particular the evolution of clustering. 
However, to achieve this goal, we first have to select the proper sample of galaxies from the collected data. 
For this purpose, we first of all need to separate the extragalactic sources from 
galactic objects (such as stars, planetary nebulae, etc.) that contaminate our data. 
This might be performed by means of follow-up observations, which are currently on-going, but they introduce additional bias in detected sources.

 AKARI data have to be categorized based on the photometric data because detailed spectroscopic follow-up observations are expensive and much more time-consuming. 
The most widely used tool in astronomy to distinguish stars and galaxies is the color-color (CC) diagram. In particular, galaxies display `redder' colors, meaning that they radiate stronger at longer wavelengths, and stars are more `blue' because they radiate strongly at shorter wavelengths (e.g. \citealt{walker}; \citealt{pollo}).
However, the methods designed up to now cannot be applied directly to 
NEP data, because they were developed for different wavebands and shallower catalogs. 
 Since different wavelengths often imply observations of different physical processes and/or different redshifts, we considered parameters obtained from several different passbands, which will enable us to distinguish sources in a multidimensional parameter space.
%
In general, classification methods are based on a pattern recognition within the data sets.
For every object we have a vector describing its characteristic features. 
We can use a mapping function, called a classifier, to transfer feature vectors into discriminant ones, which contain the likelihoods of the given object to belong to one of the considered classes.
 Classification schemes heavily depend on choosing a feature space, which should be selected in a way that different classes occupy different volumes with minimal overlapping.
When a survey is designed without a target object class (i.e., the filter sets are not specifically chosen), using unsupervised classifiers (which work without previous class information input) is a good tool to distinguish objects by, for example, using the cluster analysis (e.g. \citealt{clust}).
 This process relies on the visible features of the data.
The classification is much more obvious when we have some previous knowledge about the objects appearing in the survey. Then we can use this knowledge as an input to a supervised classifier (where we have a feature/properties template of observable objects). 
We here used the supporting vector machine (SVM) classifiers \citep{vapnik}.
SVMs are used to map input vectors non-linearly into a high dimensional parameter space and construct an optimal separating hyperplane.

 This work is organized as follows.
In Sect.~2 we give a brief description 
of the collected data together with the auxiliary survey performed by Subaru telescope \citep{iye}, which observed the NEP Deep field in optical wavelengths in filters $B, V, R, i', z'.$
 Section~3 describes the sample and parameter space selection process.
The application of the SVM method and the results are presented in Sect.~4, and its accuracy is tested by comparing our results with the separation made for optical survey of the NEP region performed by the Subaru telescope, and by preparing the flux distribution plots created for objects divided according to the established star/galaxy methods in Sect.~5. A summary and conclusions are given in Sect.~6. 

\section {The data}

The NEP Deep sky survey covers an area of 0.4 sq.$\mbox{ deg}$ around 
the North Ecliptic Pole \citep{matsuhara}.
The data were obtained by the Infra-red Camera (IRC) \citep{onaka} 
through nine near- and mid-infrared (NIR and MIR) filters, 
centered at 2$\mu$m ($N2$), 3$\mu$m ($N3$),
4$\mu$m ($N4$), 7$\mu$m ($S7$), 9$\mu$m ($S9$), 11$\mu$m 
($S11$), 15 $\mu$m ($L15$), 18 $\mu$m ($L18$), and 24 $\mu$m ($L24$) where $W$ indicates that the bandwidths are wider than the others.  
The long exposure times (from 1047 s for $N2$ filter to 261.8 s for $L24$ filter) 
mad it possible to reach very deep into this region. 
Table \ref{tab1} summarizes the survey, where $\lambda _{\rm{ref}}$ is the reference 
wavelength, $N_{\mbox{sources}}$ is the total number of detected sources 
in a specific bandpass, $\mbox{mag}_{\rm lim}$ 
 is the limiting magnitude of detected 
objects in a specific filter, and zero point stands for the magnitude zero point used 
in brightness conversion procedures. 
The point spread function (PSF) has a beam size in FWHM of $5$ arcsec, which makes AKARI's imaging superior to other infrared satellites.
The source extraction on FITS images was made with the SExtractor 
software \citep{sex}.
 A source was assumed to be detected if it had a minimum of five contiguous pixels above 
1.65 times the RMS fluctuations. Instead of allowing the program to 
estimate the background, weight maps were used. 
Photometry was carried out using SExtractor's MAGAUTO variable elliptical 
aperture with these aperture parameters: the Kron factor and the minimum radius were set to 
2.5 and 3.5. 
The magnitude zero points were derived from observations of standard 
stars \citep{tan} and were used to convert counts to magnitude by the 
photometry program.
The number of sources detected in individual filters differs significantly from each other: 
far more sources are detected in the near-infrared than in the mid-infrared.
The photometry resulted in obtaining a catalog depth of $26.86\; \mbox{mag}$ at 2.4 $\mu$m ($N2$ filter). 
The results of this procedure were downloaded from the official AKARI 
Researchers Web Page\footnote[1]{http://www.ir.isas.jaxa.jp/ASTRO-F/Observation/, however, we have performed independent photometry measurements by} SExtractor, and the parameters obtained from this run were 
used in the subsequent analysis, after confirming that the basic results 
were consistent with the original catalogs. 

\begin{table}[ht]
\caption{Properties of the NEP Deep survey based on \citealt{lor} and \citealt{wada}.}
\label{prop}
\begin{center}
\begin{tabular}{llcrrr}
\hline\hline
 Band & $\lambda_{\rm ref}$ & $N_{\rm sources}$ & $\mbox{Mag}_{\rm lim}$ & Zero point & Exp.\ {\emph t} [s] \\ \hline 
$N2$&2.4&23325&26.86&24.82&1047\\
$N3$&3.2&26180&25.95&25.02&1047\\
$N4$&4.1&26332&25.03&25.32&981.8\\
$S7$&7.0&8650&23.28&23.96&245.5\\
$S9W$&9.0&8516&22.24&24.51&212.7\\
$S11$&11.0&8769&22.93&24.19&229.1\\
$L15$&15.0&10611& 24.56&23.57&278.2\\
$L18W$&18.0&10782&23.50&23.83&278.2\\
$L24$&24.0&5704&24.59&22.28&261.8\\\hline
\end{tabular}
\end{center}
\label{tab1}
\end{table}
\subsection{Subaru/Suprime-cam optical auxiliary survey of the AKARI NEP-Deep field}
To prove the validity of our method for classifying sources we confirmed by observations that were not made in the infrared.
The best way to prove the efficiency of the presented star-galaxy separation method is to incorporate auxiliary multiwavelength data.
 The Subaru telescope observed the NEP-Deep region in $B,V,R,i',z'$ filters covering $\sim$ 0.25 $\rm deg^{2}$ (\citealt{imai}) in the field of view of the Suprime-cam (S-cam) \citep{subaru}, reaching limiting magnitudes of $z_{AB}=26$.
We cross-matched the optical data obtained by the Subaru telescope with the infrared catalogs, searching for counterparts within the radius of 5 arcsec, motivated 
by the PSF of images and known resolution of the detector. The possibility of any false identifications is assessed in Section~4.1. 
After integrating the optical data with infrared data we obtained a catalog consisting of 9699 sources in total, with 8768 optical counterparts for $NIR$ wavelengths and 3252 in $MIR.$ 
Below we use these data to test the performance of all presented methods of separation based solely on infrared data. 

\section{Sample and parameter selection}
The subsequent multivariate analysis was performed on a merged catalog of objects that were detected in all AKARI IRC passbands, which eliminates any possibility of including dropout objects.\\
As stated before, for a supervised method of classification we must adopt catalogs of known astronomical objects. 
Since we aim to develop a method based solely on IR data, we chose to use the stellarity parameter (hereafter $sgc$), an output classifier for objects based on the neural network output \citep{gur}, which is referred to as the CLASS STAR parameter, as a distinguishing value between the two desired classes.  

As one of the possible star/galaxy separation methods, 
$sgc$ was calculated by SExtractor \citep{sex} software for each source.
Detectors produce astronomical images with similar linear intensity 
scales with a good precision over large scales to the point where saturation 
takes place, therefore correctly sampled images can be roughly described by pixel scale, 
depth (signal-to-noise ratio at a given magnitude), and seeing. 
To provide the best possible classifier, input parameters should be independent of those characteristics of exposure.
Simple estimators in a two-dimensional space such as 
magnitude-isophotal area \citep{reid}, magnitude-peak intensity \citep{jones}, 
or magnitude-surface brightness \citep{harmon} are the simplest ways of separating 
stars from galaxies. 
However, the $sgc$ calculation uses ten parameters: eight isophotal areas (using more isophotal areas then the lowest one enables the classifier to be sensitive to dim objects), peak intensity (if the relative uncertainty of maximum intensity is high enough, the contrast between the two classes is worse), and seeing, which is used as a `control' parameter.
The network takes the isophotal areas in units of squared seeing FHWM, which ensures that there will be no need for the information about the pixel scale. 
The peak intensity is given in units of extraction threshold to remove the depth information.
To obtain an even more reliable classification outcome, which is independent of noise, image distortions, and influence of close objects, the SExtractor creators did not include any elongation measurements in the CLASS STAR computation. 
Its value varies between 0 and 1: 1 stands for 
a star-like object, and 0 for galaxy, or rather a non-stellar extended object. 
However, since the SExtractor is optimized to optical data, 
it is not obvious that this parameter would work for infrared observations.

\begin{figure}
\centering
\includegraphics[width=0.4\textwidth]{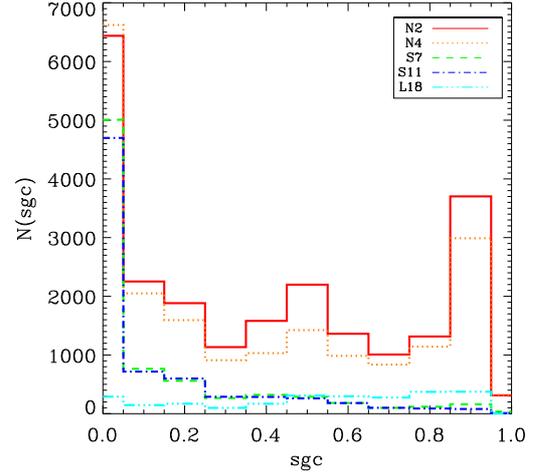} 
\caption{Stellarity parameter ($sgc$) histogram for $N2$ (solid line), $N4$ (dotted line), $S7$ (dashed line), $S11$ (dash-doted line) and $L18$ (dash-triple doted line) image. 
Here, the abscissa is the $sgc$ parameter with intervals of 0.05 and the ordinate is the number of objects in each bin.}
\label{sgc}
\end{figure}

 Fig.~\ref{sgc} represents a histogram for $sgc$ values for different wavelengths. For clarity we show distributions for five filters only, two for NIR-$N$ (solid and dotted lines)), two for MIR-$S$ (dashed and dash-dotted lines) and one for MIR-$L$ (dash-triple-dotted line).
 Evidently, for NIR filters alone, the majority of sources are unambiguously classified as extended objects (with the value 0), but a fraction of strictly star-like objects is also detected (with the $sgc$ value $\sim 1$ ) (see Fig.~\ref{sgc}).
However, the remaining $S$ and $L$-filter-based $sgc$ measurements
seem to indicate that very little or no stars are visible in these filters,
since we have only one concentration around $sgc = 0$, i.e., clearly
extended objects. 
The $sgc$ histograms for MIR wavelengths lead us to conclude that the interpretation
of this parameter at the longer wavelengths will not be useful for object classification, unlike at NIR. 
Moreover, all passbands have a local maximum around a value of $sgc = 0.5$, which means that for the applied algorithm the sources look neither clearly
extended nor clearly star-like. 
 Because of this ambiguity we decided to use the supervised approach and train the classifier based on the small but clearly determined samples of stars and galaxies instead of using the $sgc$ itself as a separator.
The training samples of star-like and extended sources are constructed in a way that objects with \emph{sgc} value in between $0 \mbox{ and }0.05$  are treated as galaxies, and $0.95\mbox{ and } 1$ are treated as stars, which resulted in obtaining training samples that consist of 825 galaxies and 532 stars.


 All possible color combinations are equally significant, therefore we kept the dimensionality of the parameter space low by choosing the infrared color indexes as follows: $N2-N3$, $N3-N4$, $N4-S7$, $S7-S11$, $S11-L15$, $L15-L18$.
Table \ref{parametryir} gives the statistical properties of the training samples, where columns list the mean values of parameters with their standard errors for selected samples of stars and galaxies. 
Clearly, that the mean values of parameters differ for different classes of sources.
 The most striking feature is the $S7-S11$ value for galaxies, which is drastically higher than for stars.
 The color index values for stars are systematically smaller. However, while in the NIR wavelength regime the differences are moderate, the discrepancy between them becomes vast when moving into longer wavelengths.
 On average, sample galaxies have the lowest flux at 7 ${\rm \mu  m}$. 
For $N4-S7$ index the difference between the two classes is marginal, and for $S7-S11$ it is most obvious.

\begin{table}[ht]
\caption{Mean values of parameters for the training samples in the multicolor space.}
\label{parametryir}
\begin{center}
\begin{tabular}{lcr}
\hline\hline
Parameter&Galaxies &Stars\\ \hline 
$N2-N3$&$-0.05\pm0.67$&$-0.32\pm0.59$\\
$N3-N4$&$-0.11\pm0.58$&$-0.53\pm0.56$ \\
$N4-S7$&$-0.15\pm0.74$&$-0.17\pm0.97$\\
$S7-S11$&$0.79\pm0.89$&$0.00\pm1.01$\\
$S11-L15$&$0.47\pm0.89$&$-0.16\pm1.08$\\
$L15-L18$&$0.03\pm0.67$&$-0.12\pm0.84$\\
\hline
\end{tabular}
\end{center}
\end{table}



\section{Support vector machines}
Support vector machines are a supervised method based on kernel algorithms (\citealt{st}) of extracting structures from data and have proven themselves to be of great use in astronomy (e.g. \citealt{wozniak}; \citealt{zz}; \citealt{hc}) due to their ability to deal with multi-dimensional data and its high accuracy. 

To train the SVM algorithm means to put in a feature vector for each object of the training example, i.e., quantities that describe the properties of a given class of objects. Therefore we maped the input data from the input space $X$ onto a feature space $H$ using a non-linear function $ \phi\!:\!X\!\rightarrow\!H.$ In the parameter space $H$ the function that will determine the boundary, which can be written as 
\begin{equation}
f(x)=\sum_{i=1}^{n}\alpha_{i}k(x,x')+b,
\end{equation}
 where $k(x,x')$ is the kernel function returning an inner product of the mapped vectors, $\alpha_i$ is a linear coefficient and $b$ is a perpendicular distance called bias, which translates it into a given direction.

 The shortest distance from the boundary to the closest points belonging to the separate classes (support vectors) is called the margin, and the algorithm searches for a hyperplane that maximizes it.
The training samples of stars and galaxies were chosen to train the 
Gaussian radial basis kernel function:
\begin{equation}
 k({\bf x},{\bf x}')=\exp(-\gamma ||{\bf x}-{\bf x}'||^2),
\label{gauss}
\end{equation}
 where $\gamma$ is the adjustable kernel width parameter, which is responsible for the curvature of the decision surface.
Since the data are not clearly separable, we added a parameter ($C$), which controls the trade-off between the misclassification and large margins.
For a more detailed description we refer the reader to \citet{hsu} or \citet{crist}.

AKARI IRC photometry provides us with nine dimensional datasets.
We reduced the number of dimensions by removing measurements in two filters: $S9$ and $L24$, since the amount of the data collected through these passbands is significantly lower than in the rest, rendering the resulting cross-matched catalog statistically insufficient.
With seven different flux measurements we built a six dimensional parameter space through using color indexes. We used two training samples containing stars and galaxies chosen according to their $sgc$ value measured in NIR to train SVM and obtain its classifier. 

 The two kernel parameters $\gamma$ and $C$ are not known beforehand and it is necessary to find the best values to obtain accurate results.
To tune these parameters for the best performance, we performed a grid-search with values from $10^{-2}$ to $10^{4}$ using a ten-fold cross-validation technique. To that end we divided the full training set into ten subsets of equal size and selected nine subsets to train the classification model and test it on the remaining subset and count the TS (true star: when an object classified as a star in the training set is classified as a star by SVM), the TG (true galaxy: when a galaxy from a training sample is classified as a galaxy by SVM), the FG (false galaxy: when a source from a star training sample is classified as a galaxy by SVM), and the FS (false star: when an object from a galaxy training sample is classified as a star by SVM). 
After concluding all iterations we summed the values and calculated the accuracy (Acc) defined as \begin{equation}
\mbox{Acc}=\frac{TS+TG}{TS+FS+TG+FG},
\label{acc}
\end{equation}
 true star rate (TSR) defined as \begin{equation}
\mbox{TSR}=\frac{TS}{TS+FG},
\label{tsr}
\end{equation} 
and true galaxy rate (TGR)  defined as \begin{equation}
\mbox{TGR}=\frac{TG}{TG+FS}.
\label{tgr}
\end{equation}
 The procedure resulted in selecting the pair ($\gamma$, $C$) equal to 1 and $10^{3},$ which provides a total accuracy of 93\%. The final results for TGR and TSR are summarized in Table \ref{confusion}.
\begin{figure}[!h]
\centering
\includegraphics[width=0.4\textwidth]{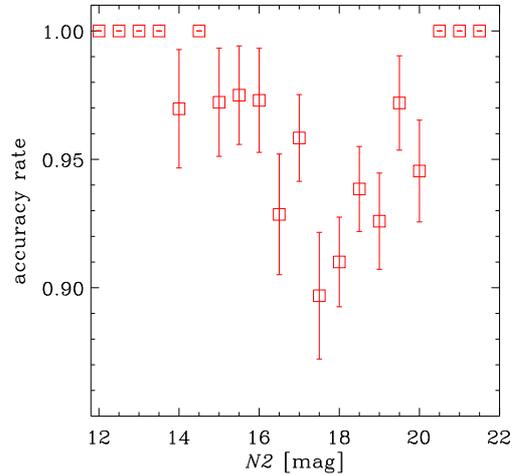} 
\caption{Accuracy rate as a function of magnitude for $N2$ passband. Error bars represent Poisson uncertainty.}
\label{accurplot}
\end{figure}

Fig.~\ref{accurplot} represents the accuracy rate (Eq. \ref{acc}) as a function of magnitude for the $N2$ filter, calculated in bins of $0.5$.
 The accuracy, though still maintaining high values, decreases with the decrease of brightness beginning from $\sim 15$ mag, 
 which may be a projection effect of bright galactic stars (which fade away in longer passbands) blocking and/or blending with the fainter extragalactic objects. 
 We explore the contamination problem below.
At the faintest end the accuracy rises again ($\sim 20.5$ mag) because in this range we have only few objects, which are mostly dim but unambiguously galaxies.

\begin{figure}[!h]
\centering
\includegraphics[width=0.4\textwidth]{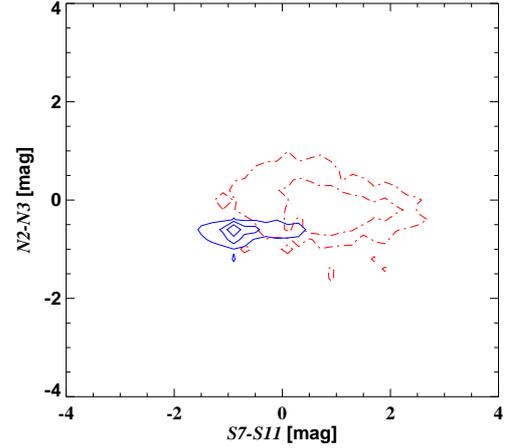} 
\caption{Projection of the SVM classification from multicolor space onto the $N2-N3$ and $S7-S11$ parameter space. Solid contours represent the occupancy zone for stars, dashed contours for galaxies.}
\label{nssvm}
\end{figure}
\begin{figure}[!h]
\centering
\includegraphics[width=0.4\textwidth]{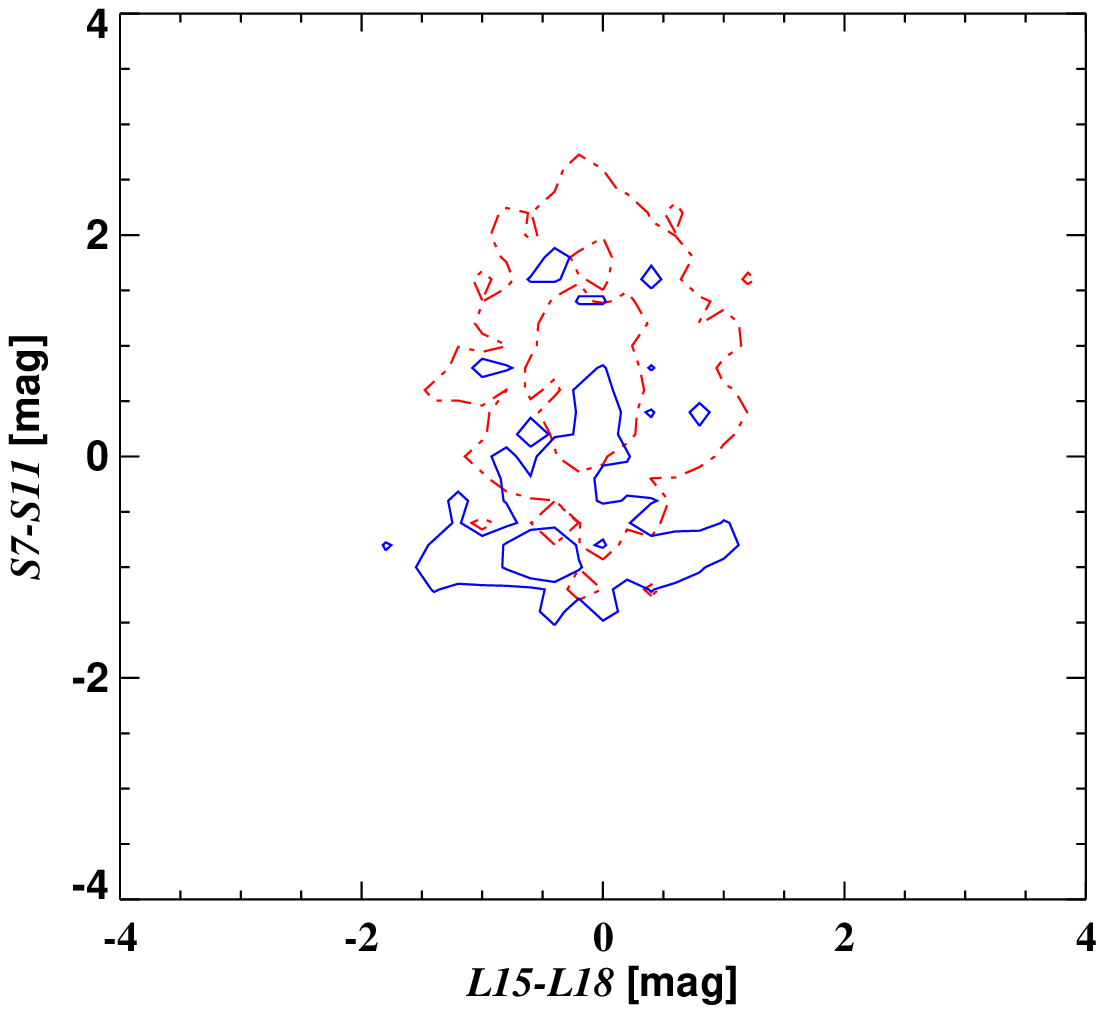} 
\caption{Projection of the SVM classification from multicolor space onto the $S7-S11$ and $L15-L18$ parameter space. Solid contours represent the occupancy zone for stars, dashed contours for galaxies.}
\label{lssvm}
\end{figure}
\begin{figure}[!h]
\centering
\includegraphics[width=0.4\textwidth]{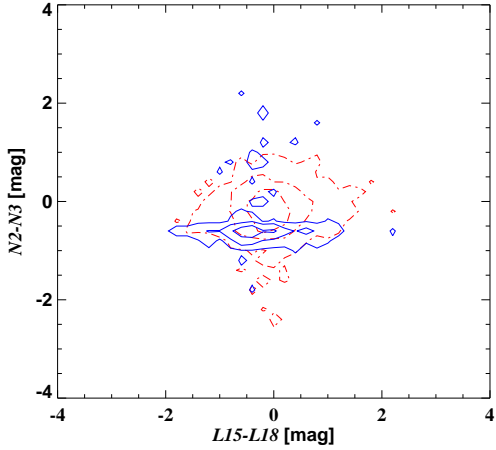} 
\caption{Projection of the SVM classification from multicolor space onto the $N2-N3$ and $L15-L18$ parameter space. Solid contours represent the occupancy zone for stars, dashed contours for galaxies.}
\label{lnsvm}
\end{figure}

Next, we applied the classifier on the whole test sample (2539 objects). 
We obtained 1657 objects classified as galaxies and 877 objects classified as stars.
As stated before, CC diagrams are the most commonly used tools for object recognition. Therefore we can assume that NIR colors are useful for selecting stars in the survey, because they radiate strongly in narrow passbands of short wavelengths, while galaxies possess much redder colors with a stronger color discrepancy, and appear as a dispersed cloud because of the variety of components comprising the spectra and their distance to the observer. To test this hypothesis  and compare it with the resulting SVM classification we projected the constructed 6D multicolor space into a standard two-color space. 
Figures \ref{nssvm}, \ref{lssvm}, and \ref{lnsvm} preset the results of the classification for the whole sample, which show the division between two classes. 
  As predicted, stars occupy compact regions of the diagrams since they have narrow emission, while galaxies tend to spread over wider range of values of color indexes.
The regions where the contours for stars and galaxies overlap indicate the projection of the class' decision boundary margin.
We fitted a linear function to the points on the 2D color space lying on the boundary hyperplane to mark the separation; the coefficients are listed in Table \ref{proste}.
\begin{table}[ht]
\caption{Performance of the trained classifier to separate stars from galaxies.} 
\label{ncn2}
\begin{center}
\begin{tabular}{lcrrr}
\hline\hline

 class& SVM star& SVM galaxy \\ \hline 
actual star&441 (TS)& 11 (FG)\\
actual galaxy& 91 (FS)& 817 (TG) \\
accuracy (\%)&98 (TSR)&90 (TGR)\\
\hline

\end{tabular}
\label{confusion}
\end{center}
\end{table}
\begin{table}[ht]
\caption{Coefficients of the linear fit to stars and galaxies lying on the 2D projection of the boundary hyperplane. }
\label{proste}
\begin{center}
\begin{tabular}{llrr}
\hline\hline
color&color &a&b\\ \hline 

$N2-N3$&$N3-N4$&$-0.14\pm0.18$&$0.08\pm0.10$\\
$N2-N3$&$N4-S7$&$0.20\pm0.12$&$0.18\pm0.09$\\
$N2-N3$&$S7-S11$&$-0.34\pm0.10$&$0.49\pm0.13$\\
$N2-N3$&$S11-L15$&$0.33\pm0.10$&$0.02\pm0.08$\\
$N2-N3$&$L15-L18$&$0.02\pm0.16$&$0.13\pm0.08$\\
$N3-N4$&$N4-S7$&$0.18\pm0.08$&$-0.29\pm0.06$ \\
$N3-N4$&$S7-S11$&$-0.04\pm0.08$&$-0.29\pm0.10$ \\
$N3-N4$&$S11-L15$&$-0.11\pm0.08$&$-0.33\pm0.06$ \\
$N3-N4$&$L15-L18$&$0.21\pm0.11$&$-0.33\pm0.06$ \\
$N4-S7$&$S7-S11$&$-0.46\pm0.10$&$0.24\pm0.13$\\
$N4-S7$&$S11-L15$&$0.22\pm0.11$&$-0.32\pm0.10$\\
$N4-S7$&$L15-L18$&$-0.08\pm0.17$&$-0.25\pm0.09$\\
$S7-S11$&$S11-L15$&$-0.59\pm0.11$&$1.25\pm0.09$\\
$S7-S11$&$L15-L18$&$0.36\pm0.18$&$1.06\pm0.10$\\
$ S11-L15$&$L15-L18$&$-0.89\pm0.15$&$0.31\pm0.08$\\
\hline
\end{tabular}
\end{center}
\end{table}

Based on the number of objects belonging to the two classes that lie within the hyperplane margin, we estimated the contamination of our samples. If an SVM classified galaxy's (or star's) position in the multicolor space has a separation boundary distance smaller than the error bar, it was treated as a possible missclassification. The contamination was estimated to be $13.16\%$ for the galaxy catalog and $9.01\%$ for the star catalog.
What is more, the missclassifications usually display a $sgc$ value of $\sim 0.5$, which confirms that these objects have to be treated with caution. 
When viewed on the FITS images, they appear to be either interacting systems and/or blended objects.
Because we aim at having a pure galaxy or stellar sample, these sources should be removed.


\section{Methodology verification}
\subsection{Subaru/Suprime-cam optical auxiliary survey of the AKARI NEP-Deep field}

 To check the validity of our star-galaxy separation methods, we created an integrated optical-infrared catalog and assigned an optical $sgc$ parameter to all test objects. 
When we transit from the optical part of the spectrum, the hot, blue stars fade away and cooler red stars come into view.
Therefore there is a possibility of a high fraction of misclassified stars from IR and optical observations. 
 Stellar NIR emission is dominated by red giants and low-mass red dwarfs.
 When we shift into MIR, the cool stars fade away and the dust-enshrouded stars emerge.
 In this regime we can observe even cooler objects, such as planets or asteroids, but considering the FOV and the depth of the survey, their contribution is not statistically significant.
 
With the same selection criteria as for the input samples, we created new training samples: 105 stars (with $0.95<sgc<1$), 226 galaxies (with $0<sgc<0.5$).
The mean values of the new training samples are summarized in Table~\ref{optmean}.

\begin{table}[ht]
\caption{Mean values of parameters for the multicolor samples based on optical $sgc$ parameter.}
\label{parametryopt}
\begin{center}
\begin{tabular}{lcr}
\hline\hline
Properties&Galaxies &Stars\\ \hline 
$N2-N3$&$-0.09\pm0.88$&$-0.49\pm0.73$\\
$N3-N4$&$0.38\pm1.12$&$0.07\pm0.94$ \\
$N4-S7$&$0.62\pm0.77$&$0.35\pm0.70$\\
$S7-S11$&$0.82\pm0.83$&$0.31\pm0.84$\\
$S11-L15$&$-0.22\pm5.44$&$0.06\pm0.52$\\
$L15-L18$&$0.49\pm5.44$&$0.13\pm0.54$\\
\hline
\end{tabular}
\end{center}
\label{optmean}
\end{table}
As expected, the farther we move into long wavelengths, the dimmer the stars become.
 Galaxy sample on the other hand have two minima at 3 ${\rm \mu m}$ and 15 ${\rm \mu m}$. 
 We compared the results with those obtained from the first classifier.
If we assume that a correctly classified object possesses the same SVM class in both IR and optical $sgc$ based classifications, then total classification has a 91\% accuracy with the TRS and TGR accuracy equal to 65\% and 96\%, respectively.
The result for the galaxy classification is very efficient. 
The efficiency is still good for stars, but it is lower than for galaxies.
 The reason is that young and hot stars that are easily detected in optical wavelengths gradually disappear when they are observed in longer wavelengths.
Therefore, their infrared counterparts could possibly be different objects, invisible in optical wavelengths due to their close proximity to bright stars. 
 This also explains the high efficiency of selecting stars based just on the IR criteria. Old, cool stars or stars with protoplanetary disks, which are hardly detectable in optical passbands, have their peak radiation in the NIR. 
Therefore it is safe to conclude that the classifier works very well for the infrared classification.

\subsection{Number counts}
 In this section we present the Euclidean normalized number counts for all considered sources and for the resulting counts for separate classes. 
We compare stellar counts with the Faint Source Model \citep{dirbe} (hereafter FSM) to assess the reliability of our results with the theoretical predictions.
 Since its primary goal was to measure the cosmic infrared background at NIR and MIR wavelengths, FSM was created as a means to remove the strong contributions of foreground emission, which originates within our Galaxy. 
At NIR wavelengths the contribution consists mainly of starlight, the majority of which can be resolved into point sources.
 However, a significant number of stars are blended into the diffuse background. 
The FSM was constructed specifically to solve this problem.
The MIR (and FIR) emission is dominated by thermal emission from dust residing in the interstellar medium and in more compact star-forming regions.  
In the wavelengths longer than 12 ${\rm \mu m}$, the faint source emission contributes less than 40\% of the observed brightness toward the inner Galaxy at low latitudes and drastically decreases (to 1\%) for higher galactic latitudes with increasing wavelength. Therefore the FSM for MIR can only follow counts in the inner Galaxy.

The measured flux in analog-to-digital units (ADU) was converted to $\mu$Jy 
by multiplying the counts by a corresponding conversion factor 
calculated for every filter \citep{lor}.
 Here we denote a flux density $ S\!_{\nu} $ at wavelength $\lambda$ $\rm{\mu m}$ as $ S\!_{ \lambda}$, but the units are [Jy].
The extragalactic Euclidean normalized differential source counts display a flat distribution at bright fluxes. If any evolution is present in the observed sample, it will be indicated by a change in the slope of counts at fainter fluxes.
If a certain population of galaxies is evolving negatively (i.e., dimming with time), the counts can be lower than the Euclidean slope. 
On the other hand, if the evolution is positive, the count slope is steeper.
At the faintest fluxes the counts will suddenly drop because of the dimming effect of cosmological redshift.

\begin{figure}[!h]
\centering
\includegraphics[width=0.4\textwidth]{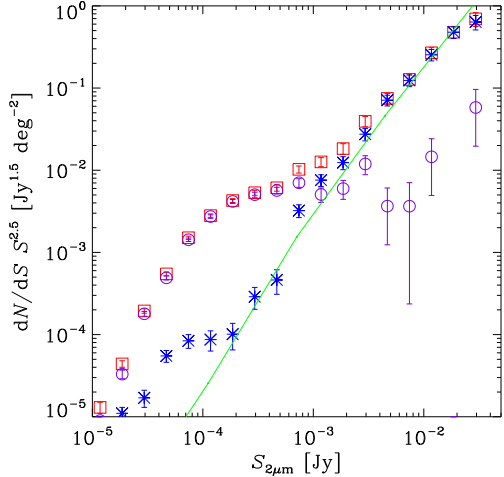} 
\caption{Euclidean normalized number counts for all objects in the sample with $N2$ fluxes represented by squares, asterisks present counts of sources classified as stellar, extragalactic counts are indicated by circles. The line presents stellar number counts predicted by the FSM. Error bars represent Poisson uncertainty in logarithmic units.}
\label{n2ff}
\end{figure}

\begin{figure}[!h]
\centering
\includegraphics[width=0.4\textwidth]{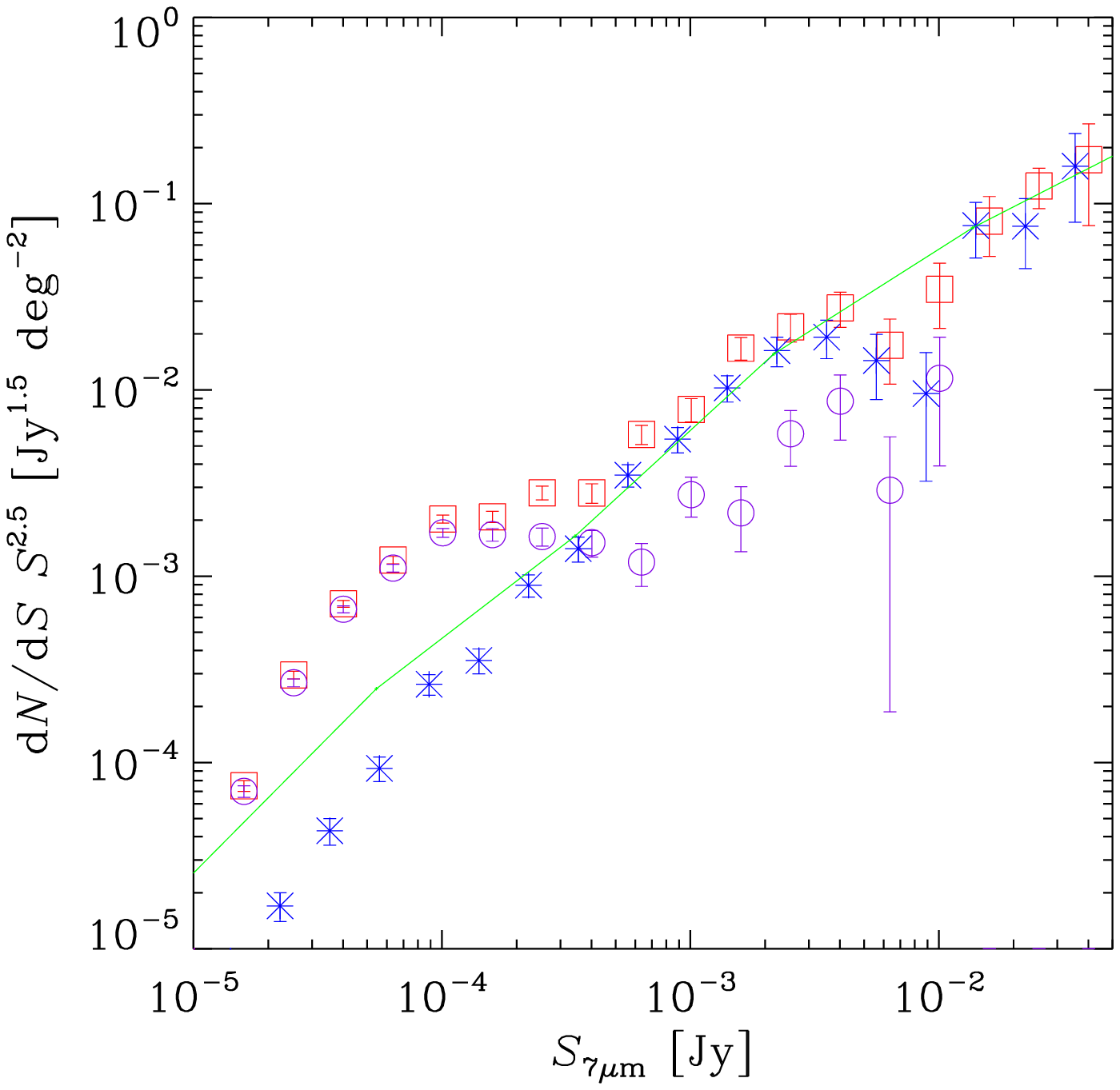} 
\caption{Euclidean normalized number counts for all objects in the sample with $S7$ fluxes represented by squares, asterisks present counts of sources classified as stellar, extragalactic counts are indicated by circles. The line presents stellar number counts predicted by the FSM. Error bars represent Poisson uncertainty in logarithmic units.}
\label{s7f}
\end{figure}

\begin{figure}[!h]
\centering
\includegraphics[width=0.4\textwidth]{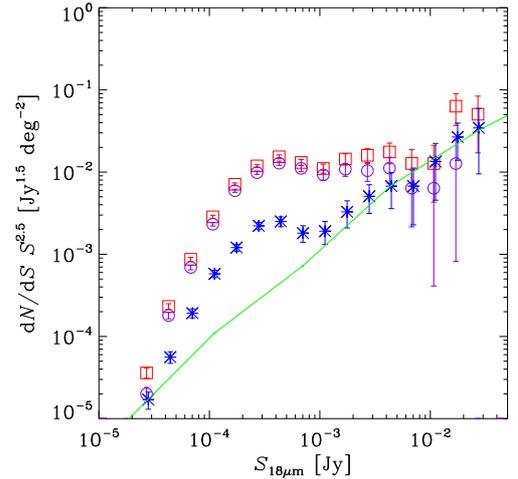} 
\caption{Euclidean normalized number counts for all objects in the sample with $L18$ fluxes represented by squares, asterisks present counts of sources classified as stellar, extragalactic counts are indicated by circles. The line presents stellar number counts predicted by the FSM. Error bars represent Poisson uncertainty in logarithmic units.}
\label{l18f}
\end{figure}

Figures \ref{n2ff}, \ref{s7f}, and \ref{l18f} present the Euclidean normalized differential number counts in 
$N2$, $ S7$, and $L18$ filters for all test objects with an assigned star or galaxy tag according to the obtained classifier.
 Squares represent total counts, asterisks present stellar counts, extragalactic counts are indicated by circles. 

In the counts for NIR wavelengths (e.g. Fig. \ref{n2ff}) we can see that the abundance of stars in the data is so high that number counts provide no distinction whatsoever.
 The raw counts show high consistency with the FSM, which proves that stars indeed dominate at the bright end of the counts. 
 The stellar counts precisely follow the theoretical predictions, and the extragalactic counts display distinctive features: a bump in counts at $N2$ filter is visible at $ S\!_{\nu}\!\sim\! 3\! \mbox{ mJy}$ together with an upturn at the brightest end.
In fainter fluxes, $ S\!_{\nu}\!<1\! \mbox{ mJy}$, the counts in these band passes slightly increase, signaling positive source evolution.
 They reach a maximum value at $ S\!_{\nu}\!\sim \!0.8\! \mbox{ mJy}$, and tail off at the faintest end, possibly because of cosmological dimming and/or catalog incompleteness. 

For MIR-$S$ bands (e.g., Fig. \ref{s7f}) the raw counts still contain a fraction of stellar sources. 
After separating stars from galaxies the expected flat distribution in extragalactic counts emerges. 
 For $ S\!_{\nu}\!<\!0.1\! \mbox{ mJy}$ the extragalactic counts have a maximum and then start to tail off.

 In MIR-$L$ bands (e.g., Fig. \ref{l18f}) the shape of total counts has changed, pronouncing evolution at fainter fluxes.
  The extragalactic counts display the Euclidean distribution to $ \!\sim \!1 \!\mbox{ mJy}$, where they start to increase, reaching a maximum value at $ \!\sim \!0.4\! \mbox{ mJy}$. Then, at $ S\!_{\nu}\!<\!0.3\! \mbox{ mJy}$, the counts begin to decrease. 
This is the effect we expect, because it is known that stars are systematically brighter than galaxies, since they are much closer to us, and this remains true also at infrared wavelengths. 
The stellar counts follow the FSM model to $ S\!_{\nu}\!\sim\! 3 \!\mbox{ mJy}$ and the shape indicates that there is a fraction of extragalactic objects classified as stars at the faintest end.
A closer look at the AKARI NEP surveys source counts was provided by \cite{pearson}. 


\section{Summary}
Measurements of the stellarity parameter carried out for near-infrared observations possess good quality for creating bimodal samples with precisely defined classes, which in this work we associated with stars and galaxies. With this knowledge alone we used these strict separating criteria to create training samples as an input to obtain the classifier, and we tested its accuracy on a test sample of objects detected in the AKARI NEP Deep field in all narrow passbands. 
We set up a six-dimensional parameter space with infrared color indexes, which have different separating values for two desired object classes.
Our training sample classifier performed well on the true classes of sources, with an accurancy of 98\% for stars and 90\% for galaxies, when considering infrared measurements alone. 
Moreover, after projecting the results into two-dimensional color spaces, we showed that the two classes overlap. However, the basic division between stars and galaxies emerges, which is consistent with the expected behavior of star/galaxy classes' occupation locus of the CC diagrams.
Nevertheless, the clear distinction is visible only in higher dimensions.
When assigning an optical value of $sgc$ to all test objects, we created a new classifier and compared the accuracy of new training sets against IR ones.
 We chose the optical $sgc$ for confirmation because SExtractor was originally designed to deal with optical data.
Our results indicate that the optical classifier works for multicolor IR data with less efficiency than the IR classifier. However, we should keep in mind that the $Subaru$ observations were carried out in a much more narrow FOV then AKARI. 
Nevertheless, the results of the comparison are still very good: 65\% of objects are classified as stars by both optical and infrared classifier, and 96\% of IR classified galaxies are pinpointed as galaxies by optical SVM.
The discrepancy for stars is probably caused by the fact that when observations move into the infrared wavelength regime, the optically bright stars start to fade away, while the optically faint objects start to emerge, overshadowing the previously bright stars.
 We also suspect for optically bright stars a chance of misclassification in the MIR catalogs more often than for other sources.
As an alternative confirmation of the accuracy of our division we created Euclidean normalized source counts for the two selected classes of objects.
 At the brightest fluxes stellar counts in all wavelengths agree well with theoretical predictions of the FSM, especially for NIR-$N$ filters, where they follow the applied model to $\sim 2 \mbox{ mJy}$.
 For MIR-$L$ the stellar contribution at MIR wavelengths is very low.
 In addition, the source counts reveal traces of positive evolution in faint fluxes in both NIR and MIR wavelengths. 
Therefore it is safe to conclude that our infrared-based classifier allows the successful selection of galactic and extragalactic objects for future analyses.

\begin{acknowledgements}
 We would like to thank the anonymous referee for providing us with very constructive and detailed comments, which greatly helped to improve and clarify the manuscript.\\
This work is based on observations with AKARI, a JAXA project with the participation of ESA. 
It made use of the NED and SIMBAD databases.
AS and AP have been supported by
the research grant of the Polish Ministry of Science Nr N N203 512938.
TTT has been supported by Program for Improvement of Research 
Environment for Young Researchers from Special Coordination Funds for 
Promoting Science and Technology, and the Grant-in-Aid for the Scientific 
Research Fund (20740105, 23340046) commissioned by the Ministry of Education, Culture, 
Sports, Science and Technology (MEXT) of Japan.
TTT and AS are partially supported from the Grand-in-Aid for the Global 
COE Program ``Quest for Fundamental Principles in the Universe: from 
Particles to the Solar System and the Cosmos'' from the MEXT.
\end{acknowledgements}

\end{document}